\def\be{\begin{equation}}
\def\ee{\end{equation}}
\def\bea{\begin{eqnarray}}
\def\eea{\end{eqnarray}}

\def\ben{\begin{enumerate}}
\def\een{\end{enumerate}}

\def\bfr{{\bf r}}

\documentclass[12pt]{iopart}
  \expandafter\let\csname equation*\endcsname\relax
  \expandafter\let\csname endequation*\endcsname\relax
\usepackage{amsmath, amssymb}

\begin{document}

\title{Problems with the Newton-Schr\"odinger Equations}

\author{C. Anastopoulos$^1$ and B. L. Hu$^2$}

\address{$^1$Department of Physics, University of Patras, 26500 Patras, Greece.}

\address{$^2$Maryland Center for Fundamental Physics and Joint Quantum Institute,\\ University of
Maryland, College Park, Maryland 20742-4111 U.S.A.}

\ead{anastop@physics.upatras.gr,blhu@umd.edu}

\date{March 30, 2014}

\begin{abstract}

We examine the origin of the Newton-Schr\"odinger equations (NSEs) that play an important role in alternative quantum theories (AQT), macroscopic quantum mechanics and gravity-induced decoherence. We show that NSEs for individual particles do not follow from general relativity (GR) plus quantum field theory (QFT). Contrary to what is commonly assumed, the NSEs are not the weak-field (WF), non-relativistic (NR) limit of the semi-classical Einstein equation (SCE) (this nomenclature is preferred over the `M\/oller-Rosenfeld equation') based on GR+QFT.  The wave-function in the NSEs makes sense only as that for a mean field describing a system of $N$ particles as $N \rightarrow \infty$, not that of a single or finite many particles. From GR+QFT the gravitational self-interaction leads to mass renormalization, not to a non-linear term in the evolution equations of some AQTs. The WF-NR limit of the gravitational interaction in GR+QFT involves no dynamics. To see the contrast, we give a derivation of the equation (i) governing the many-body wave function  from GR+QFT and (ii) for the non-relativistic limit of quantum electrodynamics (QED). They have the same structure, being linear, and very different from NSEs.  Adding to this our earlier consideration that for gravitational decoherence the master equations based on GR+QFT lead to decoherence in the energy basis and not in the position basis, despite some AQTs desiring it for the `collapse of the wave function',  we conclude that the origins and consequences of NSEs are very different, and should be clearly demarcated from those of the SCE equation, the only legitimate representative of semiclassical gravity, based on GR+QFT.

\end{abstract}

\section{Introduction and Summary}

The Newton-Schr\"odinger  equations (NSE) play a prominent role in alternative quantum theories (AQT)\cite{BassiRMP,Karol,GRWP,Diosi,Penrose}, emergent quantum mechanics \cite{AdlerCUP}, macroscopic quantum mechanics \cite{Leggett,ChenMQM,ChenNS,AHcomNS}, gravitational decoherence \cite{AHmasteq,Blen13} (such as invoked in the Diosi-Penrose models) and semiclassical gravity \cite{HuEmQM13,scg,ELE,stograCQG,HuVerLivRev}. The class of theories built upon these equations, the latest being an application of the  \textit{many-particle NSE} derived in \cite{Diosi,Penrose} to macroscopic quantum mechanics (see \cite{ChenNS,Durt} and references therein),  have also drawn increasing attention of experimentalists who often use them as the conceptual framework and technical platform for understanding the interaction of quantum matter with classical gravity and to compare their prospective laboratory results (see \cite{BassiRMP} and references therein, also \cite{Dyson}) \cite{GravDecExpts,AnalogG,VolovikHe3,CHbosenova}.

\paragraph{The NSE governing the wave function of a single particle} $\psi(\bfr,t)$ is of the form
\begin{eqnarray}
i\frac{{\partial}\psi}{{\partial t}}  = - \frac{1}{2m} \nabla^2 \psi + m^2 V_N[\psi]  \psi  \label{NS}
\end{eqnarray}
where 
$V_N(\bfr)$ is the (normalized) gravitational (Newtonian) potential given by
\be
V_N({\bf r},t) = - G \int d{\bf r'} |\psi({\bf r'},t)|^2 /|{\bf r} - {\bf r'}|.
\ee It satisfies the Poisson equation
\be \nabla^2 V_N = 4 \pi G \mu,
\ee
with the mass density
\be
\mu = m|\psi(\bfr,t)|^2
\ee
being the non-relativistic limit of energy density $\varepsilon$ corresponding to the component $T_{00}$ of the stress-energy tensor.

 The Newton-Schr\"odinger equations' admittance of spatial localization of the wavefunction makes it attractive to many AQTs who view the"collapse of the wave function" in space for macroscopic objects as a strong motivation for seeking departures from quantum mechanics. Since this brings about
  the same qualitative result as gravitationally-induced decoherence --  NSE is often attributed this added laurel \footnote{In contrast, a master equation derived   from GR and QFT predicts decoherence in the energy rather than the position basis, with negligible magnitude \cite{AHmasteq}.}.
However, the mathematical foundation and physical soundness  of the Newton-Schrodinger equations seem shaky to us.
In this paper we examine the structure of NSE in relation to general relativity (GR) and quantum field theory (QFT), the two well-tested theories governing the dynamics of classical spacetimes and quantum matter.   The viability of NSEs is usually   assumed  courtesy of their well-accepted progenitor theories. Since Newtonian gravity is the weak field (WF) limit of GR, and quantum mechanics is the nonrelativistic (NR) limit of quantum field theory, it is easy to slip into believing that NSE is a limiting case derivative of GR and QFT. However, when the weak-field (Newton) and non-relativistic (Schr\"odinger) forms are taken on face value, subtle points are ignored, leading to a class of theories that are very different from, in fact, contradicting,  the conjunction of GR and QFT. In this paper,  we  cross-examine these practices and expound the assumptions which proponents of theories based on NSE often make for stated purposes, but provide little justification.

To get a taste of this, we mention here two clear differences in the physical features and consequences between 1) the NS Equation based on Newton's gravity and Schr\"odinger's quantum mechanics, using single- or many-particle wave functions. 2) the WF-NR limit of quantum field theory in curved spacetime  where gravity is described by general relativity and matter described by quantum fields, interacting with gravity in the proper manner. After this we will describe the two approaches we took which led us to these conclusions.

\subsection{NS Equation not from GR + QFT}

A. In NSE, the {\em gravitational self-energy}  defines non-linear terms in Schr\"odinger's equation.  In comparison, in the class of AQTs proposed by Diosi \cite{Diosi}, the gravitational self-energy defines a stochastic term in the master equation.
With GR and QFT, the gravitational self-energy only contributes to mass renormalization in the weak field limit. The Newtonian interaction term induces a divergent self-energy contribution to the single-particle Hamiltonian. It does not induce any nonlinear term in the evolution of single-particle wave-functions.


B. The {\em  single-particle `wave function' in the Newton-Schr\"odinger equation} $\chi({\bf r})$
appears as a result of making a Hartree approximation for $N$ particle states as $N \rightarrow \infty$. Consider the ansatz $|\Psi\rangle = |\chi\rangle \otimes |\chi \rangle \ldots \otimes|\chi \rangle$ for a $N$-particle system. At the limit $N \rightarrow \infty$, the generation of particle correlations in time is suppressed and one gets an equation which reduces to the NS equation for $\chi$ \cite{Alicki,AdlerTDG}. However, in the Hartree approximation, $\chi({\bf r})$ is \textit{not} the wave-function  of a single particle, but a \textit{collective variable} that describes a system of $N$ particles under a mean field approximation\footnote{Note that it is long known \cite{HarHor} that the semi-classical Einstein equations corresponds to the large-N limit of N component quantum fields living in a curved spacetime. See also \cite{RVlargeN} for the next-to-leading order large N expansion giving rise to the Einstein-Langevin equation in stochastic gravity theory.}.

This shows what could go wrong if one stays at the  restricted  level of particle wave-functions (rather than the more basic and accurate level of QFT) in exploring the interaction of quantum matter with classical gravity. The one-particle, or the many-particle, NS equation \cite{ChenNS} is not a  physical representation of how quantum matter is coupled to classical gravity or how it is accommodated in curved spacetimes.

Point A above explains why nonlinearity does not arise in a proper QFT treatment. Point B indicates that the interaction of quantum matter with classical gravity is only meaningful if the matter degrees of freedom are fundamentally  described in terms of quantum fields. A coupling of gravity and matter through the single-particle wave functions in quantum mechanics is like treating them implicitly  as {\em classical fields}. This mars their probabilistic role in quantum theory. Like all non-linear modifications to Schr\"odinger's equation, it is not clear how to interpret such wave-functions when considering probabilities in statistical ensembles.
Subtle differences, such as the one between a quantum mechanical versus a QFT treatment of quantum matter in the presence of gravity,  result in markedly varied consequences.




The above observations came from analysis we performed via two routes:
1) Taking the non-relativistic limit of the semiclassical Einstein equation (SCE), 
the central equation of relativistic semiclassical gravity   \footnote{There are four levels of semiclassical gravity (SCG) theories \cite{HuEmQM13} and, to avoid confusion when discussing issues, one needs to specify which level of SCG one refers to. Our suggestion is to use the two most developed levels \cite{HuVerLivRev} which we refer to as "relativistic semiclassical gravity" here.}, a fully covariant theory based on GR+QFT \cite{BirDav,Fulling,Wald,ParTom} with self-consistent backreaction of quantum matter on the spacetime dynamics (for discussions of the criterion and range of its validity, see \cite{FlaWal,HRValidity}).   2) Working out from first principles  a model with matter described by a  scalar field interacting with weak gravity (see \cite{AHmasteq}), solve the constraint, canonically quantize the system,  then take the nonrelativistic limit. This procedure is  analogous to the derivation of the non-relativistic limit of quantum electrodynamics (QED). The equations obtained in both cases have the same structure, ostensibly linear, and very different from NSEs.


\subsection{Non-relativistic weak field limit of SCE equation}

The semiclassical Einstein equation \footnote{We prefer calling it the  semiclassical Einstein  equation over the `M\/oller-Rosenfeld equation' \cite{MolRos} because after all it is Einstein's equation, albeit with a quantum matter source.} is of the form
\be G_{\mu \nu} = 8 \pi G \langle \Psi|\hat{T}_{\mu \nu}|\Psi \rangle, \label{sce} \ee
where $\langle \hat{T}_{\mu \nu} \rangle$ is the expectation value of the stress energy density operator $\hat{T}_{\mu \nu}$ with respect to a given   quantum state $| \Psi \rangle$ of the field. One usually employs the Heisenberg picture in the spacetime argument of the operator $\hat{T}^{\mu \nu}$; the state $|\Psi\rangle$ is constant in time.

In the weak field limit, the spacetime metric has the form $ds^2 = (1 - 2 V)dt^2 - d{\bf r}^2$.  The semi-classical Einstein equation becomes
\begin{eqnarray}
\nabla^2 V = 4 \pi G \langle \hat{\varepsilon}\rangle, \label{semi}
\end{eqnarray}
where $\hat{\varepsilon} = \hat{T}_{00}$ is the energy density operator. The Newtonian potential is not a dynamical object in GR, just like the electric potential is not dynamical in QED, but it is expressed in terms of dynamical variables through first-class constraints.

Eq. (\ref{semi}) can be solved to yield
\begin{eqnarray}
V(\bfr) = - G \int d{\bf r'} \frac{\langle \Psi|\hat \varepsilon(\bfr')|\Psi\rangle}{|{\bf r} - {\bf r'}|}. \label{ehat}
\end{eqnarray}
The expectation value of the stress energy tensor has ultraviolet divergences and needs to be regularized. Such regularization procedures were investigated in the mid-70's with well known results (see, e.g., \cite{BirDav}).

In the nonrelativistic limit,
$\hat \varepsilon({\bf r'})$ becomes $\hat{\mu}_{reg}(\bfr) $, the regularized mass density operator
The evolution of the quantum field is described by an `effective Hamiltonian'
\begin{eqnarray}
\hat{H} = - \frac{1}{2m} \int d \bfr \hat{\psi}^{\dagger}({\bfr}) \nabla^2 \hat{\psi}({\bfr}) 
- G \int d \bfr d \bfr' \hat{\mu}_{reg}({\bfr})\frac{\langle \Psi|\hat{\mu}_{reg}({\bfr'})|\Psi\rangle}{|{\bfr} - {\bfr'}|}. \label{nrSCE}
\end{eqnarray}
where $\hat{\psi}(\bfr), \hat{\psi}^{\dagger}(\bfr)$ are respectively the non-relativistic field annihilation and  creation operators   expressed in the position basis---for the precise definition, see Eq.  (\ref{psii}).

One could assume that the relevant field states $|\Psi\rangle$ correspond to a single particle  and derive the NS equation for a single particle from Schr\"odinger's equation associated to the Hamiltonian (\ref{nrSCE}). But such a procedure violates the way quantum matter fields are supposed to be coupled to gravity in Eq. (\ref{sce}). The SCE equation is meaningful as an approximation to a more fundamental quantum theory of gravity only in the mean field limit, with the expectation values of matter fields acting as source, and is viable only for states $|\Psi\rangle$ for which the mean-field approximation is valid. Single-particle (or even few-particle) states do not belong to this class.

The specific   procedure  leading one from SCE to a NS equation in the description above  is the treatment of $ m |\phi(\bfr)|^2 $ as a mass density for a single particle  described by the wave-function $\phi(\bfr)$.
The problem with this procedure is that the mass density is in fact an observable (rather than a part of the wave-function), and it corresponds to an operator $\hat{\mu}_{reg}({\bf r}) = m \hat{\psi}^{\dagger}({\bf r}) \hat{\psi}({\bf r})$ in the QFT Hilbert space.

The   field state
\begin{eqnarray}
|\phi\rangle = \int \hat{\psi}^{\dagger}(\bfr) \phi(\bfr)|0\rangle, \label{1part}
 \end{eqnarray}
 where $|0\rangle$ is the vacuum, describes a single particle. For this state, the expectation value  $\langle \phi|\hat{\mu}_{reg}({\bf r})|\phi\rangle $ indeed coincides with $ m |\phi(\bfr)|^2 $. However, the substitution of an operator with its mean value is a good approximation {\em only if the system is presupposed to behave classically}. In the context of the SCE equation, such an approximation is meaningful only at the mean-field description of a many-particle system. When considering a single particle, the mass-density ought to be treated as an operator in the evolution equations.

This misstep leads one to the consequences A and B, described in the beginning of Sec. 1.1. Starting from GR and QFT, one sees no nonlinearity in the dynamical equations for the matter field. One- or many- particle NSEs is not derivable from GR and QFT \cite{AHcomNS}.

\subsection{Perils of single particle wave function}

 In Sec. 1.2, we described the procedure of starting from the SCE and identifying the step which misleads one to the NSE for single or finitely many particles. We have also carried out an explicit calculation following a  procedure   detailed in \cite{AHmasteq}, namely,  consider classical matter interacting with weak gravity (perturbations off the Minkowski metric) solving the constraints, quantizing, and then  taking the non-relativistic limit.  

The result is a Schr\"odinger's equation for the state $|\Psi\rangle$ associated to the quantum field
\begin{eqnarray}
i \frac{\partial |\Psi \rangle} {\partial t}  = \hat{H} |\Psi \rangle, \label{eqsch}
\end{eqnarray}
where the QFT Hamiltonian is
\begin{eqnarray}
\hat{H} = - \frac{1}{2m} \int d{\bf r}
\hat{\psi}^{\dagger}({\bf r}) \nabla^2 \hat{\psi}(\bfr) 
- G \int  \int d{\bf r} d{\bf r'} \frac{(\hat{\psi}^{\dagger}\hat{\psi})({\bf r}) (\hat{\psi}^{\dagger}\hat{\psi})({\bf r'})}{|{\bf r} - {\bf r'}|}, \label{ham}
\end{eqnarray}
expressed in  terms of the non-relativistic field operators $\hat{\psi}(\bfr), \hat{\psi}^{\dagger}(\bfr)$.

The electromagnetic analog of Eq. (\ref{ham}) with the Coulomb potential replacing the gravitational potential is widely used in condensed matter physics.

Let us see what Eq. (\ref{ham}) looks like when projected down to single-particle states of the form (\ref{1part}).
The matrix elements of the operator (\ref{ham}) with respect to a pair of single-particle states define the single-particle Hamiltonian:
\begin{eqnarray}
\langle \phi_2|\hat{H}|\phi_1 \rangle = - \frac{1}{2m} \int d{\bf r} \phi_2^*({\bf r})\nabla^2 \phi_1({\bf r}) 
 - G \int d {\bf r} d{\bf r'} \frac{\phi_2^*({\bf r'})  \phi_1({\bf r}) \delta ({\bf r} - {\bf r'})}{|{\bf r} - {\bf r'}|}. \label{singpar}
\end{eqnarray}
The second term on the right-hand-side of Eq. (\ref{singpar}) is an infinite constant added to the single-particle Hamiltonian, i.e.,  a divergent
 self-energy contribution. Eq. (\ref{ham}) does not induce any nonlinear term in the evolution equation.

  In our opinion, the correct description of quantum matter interacting
with classical gravity is if the matter degrees of freedom are described in terms of
quantum fields, not in terms of single-particle wave functions whose dynamics NSEs purport to describe.  One can obtain a single- or N-particle description by projecting the end results of quantum matter fields interacting with classical gravity onto the 1 or N particle sectors. We have explicitly provided these equations in this paper, which are ostensibly different from the NSEs for single- or N- particles obtained from using the single- or N- particle wave functions  {\it ab initio} in the Schr\"odinger equation. (On the issue of a quantum field description versus single quantum particle description of quantum matter interacting with a classical gravitational field,  see also \cite{kiefer,lammer}.)  We assert that the only valid theory for the interaction of quantum matter with classical gravity based on the two well-known and well-tested theories GR+QFT in their respective validity domains,  is (relativistic) semiclassical gravity which offers a mean field description, or stochastic gravity, with the inclusion of quantum matter fluctuations \cite{HuEmQM13}.

Our main conclusion is that NSEs do not follow from general relativity  plus quantum field theory. Thus,  all theories based on or making use of NSEs assume some unknown physics which need be justified and verified. This  may be the attitude taken by some proponents of AQTs, that their theories are beyond existing physics. Our modest goal here is to provide an explicit theoretical platform, built purely from GR and QFT, so that all proposers of AQTs can bring their favorite theories to compare with, to explain and better justify their logical reasons for existence.

This paper is organized as follows: In Sec. 2, we briefly describe the well-known derivation of the non-relativistic limit in QFT, in order to make explicit the points referred to above.  We include the definition of the regularized mass-density operator. In Sec. 3, we sketch our model for gravity-matter coupling and show the derivations leading to the Hamiltonian (\ref{ham}) above  or the equivalent Eq. (\ref{field1}) below. Details are contained in Appendix A. We write down the Hamiltonian for one particle, two particle and the mean field. From these expressions one can see explicitly the differences with the NSEs.   In Sec. 4, following the same procedure, we work out the analogous problem in QED;  the non-relativistic limit of QED is a well-accepted theory used in condensed matter physics. We draw our conclusions in Sec. 5. An alternative derivation using a different procedure, that of first taking the Newtonian limit, then quantizing the system and then solving the constraint, gives the same result for the WF-NR limit as the fully relativistic treatment. An outline of this alternative is given in Appendix B.

 \section{Nonrelativistic limit of Quantum Field Theory}

In this section, we briefly present the derivation of the non-relativistic limit in a scalar QFT. We  also define  the regularized mass-density operator.

Consider a scalar quantum field $\hat{\phi}(\bfr)$ and its conjugate momentum $\hat{\pi}(\bfr)$ expressed in terms of the creation and annihilation operators $\hat{a}_{\bf k}$ and $\hat{a}^{\dagger}_{\bf k}$
\begin{eqnarray}
\hat{\phi}(\bfr) = \int \frac{d^3k}{(2\pi)^3 \sqrt{2 \omega_{\bf k}}} \left[ \hat{a}_{\bf k} e^{i {\bf k}\cdot \bfr} + \hat{a}^{\dagger}_{\bf k} e^{- i {\bf k}\cdot \bfr}\right] \label{fieldq1} \\
\hat{\pi}(\bfr) = i \int \frac{d^3k}{(2\pi)^3 } \sqrt{\frac{ \omega_{\bf k}}{2}}\left[ - \hat{a}_{\bf k} e^{i {\bf k}\cdot \bfr} + \hat{a}^{\dagger}_{\bf k} e^{- i {\bf k}\cdot \bfr}\right]. \label{fieldq2}
\end{eqnarray}
For a free field, the Hamiltonian operator is
\begin{eqnarray}
\hat{H} = \int \frac{d^3k}{(2\pi)^3 }  \omega_{\bf k} \hat{a}^{\dagger}_{\bf k} \hat{a}_{\bf k},
\end{eqnarray}
where $\omega_{\bf k} = \sqrt{{\bf k}^2 + m^2}$.

In the non-relativistic approximation, we define the fields
\begin{eqnarray}
\hat{\psi}(\bfr) = \int   \frac{d^3k}{(2\pi)^3 } \hat{a}_{\bf k} e^{i {\bf k}\cdot \bfr} , \hspace{1cm}
\hat{\psi}^{\dagger}(\bfr) = \int \frac{d^3k}{(2\pi)^3 } \hat{a}^{\dagger}_{\bf k} e^{- i {\bf k}\cdot \bfr}, \label{psii}
\end{eqnarray}
and we approximate
\begin{eqnarray}
\hat{\phi}(\bfr) = \frac{1}{\sqrt{2m}} \left[ \hat{\psi}(\bfr) + \hat{\psi}^{\dagger}(\bfr) \right],   \hspace{1cm}
\hat{\pi}(\bfr) = - i \sqrt{\frac{m}{2}} \left[ \hat{\psi}(\bfr) - \hat{\psi}^{\dagger}(\bfr) \right]. \label{newt}
\end{eqnarray}
The Hamiltonian then becomes
\begin{eqnarray}
\hat{H} = m \int d \bfr \hat{\psi}^{\dagger}(\bfr) \hat{\psi}(\bfr)  - \frac{1}{2m}\int d \bfr \hat{\psi}^{\dagger}(\bfr) \nabla^2 \hat{\psi}(\bfr). \label{hamnr}
\end{eqnarray}
We will denote the second term in Eq.(\ref{hamnr}) as $\hat{H}_0$ because it corresponds to the Hamiltonian for $N$ non-relativistic particles. The particle-number operator $\hat{N}$ is
\begin{eqnarray}
\hat{N} = \int d \bfr  \hat{\psi}^{\dagger}(\bfr) \hat{\psi}(\bfr). \label{numop}
 \end{eqnarray}

This suggests that $ m \hat{\psi}^{\dagger}(\bfr) \hat{\psi}(\bfr)$ can be identified as the mass-density \textit{operator}. However, the expression  $\hat{\psi}^{\dagger}(\bfr) \hat{\psi}(\bfr)$ does not correspond to a  well-defined self-adjoint operator.

We define a  {\em regularized} mass density operator
\begin{eqnarray}
\hat{\mu}_{reg}(\bfr) = m \int d\bfr' \varsigma_{\sigma}(\bfr' -\bfr) \hat{\psi}^{\dagger}({\bfr'}) \hat{\psi}({\bfr'}), \label{massden}
\end{eqnarray}
using a smearing function $\varsigma_{\sigma}(\bfr)$   that satisfies the conditions
\begin{enumerate}
\item $\varsigma_{\sigma}(\bfr) \geq 0$.
\item  $ \lim_{\sigma \rightarrow 0} \varsigma_{\sigma}(\bfr) = \delta^3(\bfr)$.
\item $\int d^3x \varsigma_{\sigma}(\bfr) = 1$.
\end{enumerate}
A convenient choice for $\varsigma_{\sigma}$ is the Gaussian function
\begin{eqnarray}
\varsigma_{\sigma}(\bfr) = (2 \pi \sigma^2)^{-3/2} e^{- \frac{\bfr^2}{2\sigma^2}}. \label{gaussian}
\end{eqnarray}


\section{Matter Field interacting with gravity in the weak-field nonrelativistic limit}

In what follows, we show an explicit derivation of (\ref{ham}) following the procedures used in \cite{AHmasteq}. We will not dwell on the open quantum system aspects therein, whereby a master equation for gravitational decoherence is derived.

 \subsection{Derivation of the field Hamiltonian}
Consider a classical   scalar field $\phi$ of mass $m$ describing the matter degrees of freedom and its interaction
 with a gravitational field.  The action for this system is
\begin{eqnarray}
S[g,\phi] = \frac{1}{\kappa}\int  d^4x \sqrt{-g} R +  \int d^4x \sqrt{-g}
\left(-\frac{1}{2} g^{\mu \nu} \nabla_{\mu} \phi \nabla_{\nu}\phi -
\frac{1}{2}{ m^2} \phi^2\right), \label{Scov}
\end{eqnarray}
where $\nabla_{\mu}$ is the covariant derivative defined on a background spacetime with Lorentzian metric  $g_{\mu\nu}$, $R$ is the spacetime's Ricci scalar, $g$ is the determinant of the metric and $\kappa = 8 \pi G$; $G$ is Newton's gravitational constant.

In Appendix A, we summarize the 3+1 treatment of the action (\ref{Scov}) in the weak gravity limit. We consider  linearized perturbations of the metric around the Minkowski spacetime, we implement the Legendre transform to pass on to the Hamiltonian description, then we perform the constraint analysis.

The end result is the Hamiltonian
\begin{eqnarray}
H = \frac{1}{2}\int d \bfr ( \pi^2 +(\nabla \phi)^2 + m^2 \phi^2) + H_{TT} - \frac{\kappa}{2} \int d \bfr \bar{\gamma}^{ij} \mathfrak{t}_{ij} \nonumber \\
-\frac{\kappa}{8 \pi}  \int d \bfr d \bfr'  \frac{\epsilon(\bfr)\epsilon(\bfr')}{ |{\bf r} - {\bf r'}|} +H'_{int} \label{hfin0}
\end{eqnarray}
where $\pi$ is the conjugate momentum to the scalar field $\phi$, $\bar{\gamma}^{ij}$
are the transverse-traceless metric perturbations, $H_{TT}$ is the self-Hamiltonian for the transverse-traceless perturbations, $\mathfrak{t}_{ij} $ is the spatial components of the field's stress-energy tensor, $\epsilon$ is the energy density
\begin{eqnarray}
\epsilon(\bfr) = \frac{1}{2}( \pi^2 +(\nabla \phi)^2 + m^2 \phi^2), \label{eden}
\end{eqnarray}
and $H'_{int}$ refers to other interaction terms that are negligible in the non-relativistic limit.

The Hamiltonian (\ref{hfin0}) follows from solving the constraints of linearized general relativity, Eqs. (\ref{scalcon}) and (\ref{veccon}),  at the classical level. The term involving the energy density $\epsilon$ is the only one that survives in the non-relativistic limit, because it contains the mass density $\mu(\bfr)$, which is the only source of the gravitational field in the Newtonian regime.

We then proceed to canonically quantize the system. In particular, we substitute the classical fields $\phi(x), \pi(x)$ with the quantum operators (\ref{fieldq1}---\ref{fieldq2}), and similarly for the 3-metric $\hat{h}_{ij}$ and its conjugate momentum $\hat{\Pi}^{ij}$. Having quantized the fields $\hat{\phi}$ and $\hat{\pi}$, a regularized expression for the energy density $ \epsilon(\bfr)$ is straightforwardly defined as the quantum version of Eq. (\ref{eden}).
The resulting field theory is well defined at the  tree level.

This procedure follows the prescription of {\em reduced state space quantization}. An alternative procedure is to quantize the system prior to the imposition of the constraints; this is the essence of Dirac quantization. In general, the  two procedures  produce different results. But it turns out that they lead to the same result in the non-relativistic limit, mainly because the scalar constraint of general relativity, Eq. (\ref{scalcon}), becomes very simple. The alternative derivation is sketched in Appendix B. Either procedure is standard for the quantization of constrained systems. The one we present in this section corresponds to the standard derivation of the non-relativistic limit of QED in atomic and many-body systems as we will see in the next section.

In Ref. \cite{AHmasteq}, we quantized both the scalar field $\phi$ and the gravitational perturbations $\bar{\gamma}_{ij}$ in Eq. (\ref{hfin0}), we derived a master equation for the quantized matter field $\hat{\phi}$ and then took the non-relativistic particle limit. The emphasis there was on possible  decoherence effects due to gravitational perturbations -- see also \cite{An96} and \cite{Blen13}.

Here, we explore a different regime and we ignore the effect of the gravitational perturbations. Thus, we need not consider the $H_{TT}$ term and the term coupling the perturbations to the spatial components of the stress-energy tensor in Eq. (\ref{hfin0}).

We take the non-relativistic limit as in Eq. (\ref{hamnr}) for the free field terms in Eq. (\ref{hfin0}). Classically, the energy density $\epsilon(\bfr)$ coincides with the mass density $\mu(\bfr)$ in the non-relativistic limit. In the quantum description, the regularized operator    $\hat{\epsilon}(\bfr)$ for the energy density, i.e., the quantized version of Eq. (\ref{eden}), is substituted by the regularized  mass density operator $\hat{\mu}_{reg}(\bfr)$, Eq. (\ref{massden}).

The result is the Hamiltonian operator
\begin{eqnarray}
\hat{H} = m \hat{N}  - \frac{1}{2m} \int d{\bf r}
\hat{\psi}^{\dagger}({\bf r}) \nabla^2 \hat{\psi}({\bf r}) - G \int d\bfr d \bfr' \frac{\hat{\mu}_{reg}({\bf r}) \hat{\mu}_{reg}({\bf r'})}{|{\bf r} - {\bf r'}|}. \label{field1}
\end{eqnarray}
Eq. (\ref{field1}) is the main results in this approach. Restricting the Hamiltonian to the $N$-particle subspace, we obtain the effective gravitational dynamics of $N$ particles.

\subsection{One-, two-particle states and mean field limit}
\bigskip

{\em One particle.} We first consider a single particle state
\begin{eqnarray}
| \phi \rangle = \int d \bfr \hat{\psi}^{\dagger}({\bf r}) \phi({\bf r}) | 0 \rangle, \label{1par}
\end{eqnarray}
where $\phi({\bf r})$ is the single-particle wave-function.

The matrix elements of the operator (\ref{field1}) on the single-particle states are
\begin{eqnarray}
\langle \phi_2|\hat{H}|\phi_1 \rangle = - \frac{\hbar^2}{2m} \int d \bfr \phi_2^*({\bf r})\nabla^2 \phi_1({\bf r})
+  \delta m_{\sigma} \int d \bfr \phi_2^*({\bf r}) \phi_1({\bf r}),
\end{eqnarray}
where
\begin{eqnarray}
\delta m_{\sigma} = - \frac{G m^2}{\sqrt{\pi \sigma^2}}.
\end{eqnarray}
Hence, the Hamiltonian operator in the one-particle subspace is
\begin{eqnarray}
\hat{H} = m_{ren} \hat{1}+ \frac{\hat{p}^2}{2m} ,
\end{eqnarray}
where $\delta m_{\sigma}$ has been absorbed into mass renormalization $m_{ren} = m + \delta m_{\sigma}$.

Thus,  the Newtonian interaction term at the field level induces a divergent self-energy contribution to the single-particle Hamiltonian. It does \textit{not} induce non-linear term with respect to the particle wave functions. In particular,  the NS equation {\em is not} the evolution equation for the single-particle wave function.

\bigskip

{\em Two particles.} Next, we consider a 2-particle state
\begin{eqnarray}
|\phi_1, \phi_2 \rangle = \frac{1}{\sqrt{2}} \int d \bfr_1 d \bfr_2 \phi_1({\bf r}_1) \phi_2({\bf r}_2) \hat{\psi}^{\dagger}({\bf r}_1) \hat{\psi}^{\dagger}({\bf r}_2)|0 \rangle.
\end{eqnarray}
Let us denote by $\hat{H}_I$ the interaction term in the Hamiltonian (\ref{field1}), that is,
\begin{eqnarray}
\hat{H}_I = - G \int d^3{\bf r} d^3 {\bf r'} \frac{\hat{\mu}_{reg}({\bf r}) \hat{\mu}_{reg}({\bf r'})}{|{\bf r} - {\bf r'}|}.
\end{eqnarray}
The corresponding matrix elements of the Hamiltonian (\ref{field1}) are
\begin{eqnarray}
\langle \chi_1, \chi_2|\hat{H}_I|\phi_1, \phi_2 \rangle = 2 \delta m_{\sigma} \langle \chi_1, \chi_2|\phi_1, \phi_2 \rangle \nonumber \\
 - G m^2 \int d \bfr d \bfr' F_{\sigma}(|{\bf r} - {\bf r}'|) \left[ (\bar{\chi}_1\phi_1)({\bf r}) (\bar{\chi}_2\phi_2) ({\bf r}') + (\bar{\chi}_1\phi_2)({\bf r}) (\bar{\chi}_2\phi_1) ({\bf r}') \right],
\end{eqnarray}
where
\begin{eqnarray}
F_{\sigma}(r) = \frac{1}{r} \mbox{Erf}\left(\frac{r}{2 \sigma}\right),
\end{eqnarray}
is a regularized version of the Newtonian potential. We note that as $\sigma \rightarrow 0$, $F_{\sigma}(r) \rightarrow 1/r$.

Thus, the Hamiltonian on the 2-particle subspace is
\begin{eqnarray}
\hat{H} = 2 m_{ren}\hat{1} + \frac{\hat{{\bf p}}^2_1}{2m} + \frac{\hat{{\bf p}}^2_2}{2m} -  \frac{G m^2}{|\hat{{\bf r}}_1 - \hat{\bf r}_2|},
\end{eqnarray}
 where the self-interaction term $2 \delta m_{\sigma}$ has been consistently absorbed in the mass renormalization. Again,   no NS equation appears.

\bigskip

{\em The mean-field limit.}
In the $N$-particle subspace, the Hamiltonian becomes (modulo the renormalized mass term)
\begin{eqnarray}
\hat{H} = \sum_{i = 1}^n \frac{\hat{{\bf p}}^2_i}{2m} -\sum_{i \neq j} \sum_j  \frac{G m^2}{|\hat{\bf r}_i - \hat{\bf r}_j|}.
\end{eqnarray}

We consider $N$-particle states of the form
\begin{eqnarray}
|\Psi\rangle = |\chi\rangle \otimes |\chi \rangle \ldots \otimes|\chi \rangle := \otimes_{i = 1}^N |\chi\rangle
\end{eqnarray}
   where $\chi({\bf r})$ is a single-particle wave-function. Then, the following theorem applies \cite{meanfield1, meanfield2}
\begin{eqnarray}
\lim_{N \rightarrow \infty} e^{-i \hat{H}t} \otimes_{i = 1}^N |\chi\rangle = \otimes_{i = 1}^N |\chi(t) \rangle
\end{eqnarray}
   where $\chi(\bfr, t)$ satisfies the Newton-Schr\"odinger equation.
      However, in this approximation, $\chi({\bf r}, t)$ is \textit{not} the wave-function of a single particle, but a \textit{collective variable} that describes a system of $N$ particles.


\section{The electromagnetic analogue: nonrelativistic limit of QED }

In this section we consider the analogue electromagnetic (EM) system, namely, scalar QED, which describes the interaction between a charged particle represented here by a  complex scalar field $\phi$ and an electromagnetic field with vector potential $A_{\mu}$. Of course, there exist basic differences between gravity and EM, such as the nonlinearity of the former but the linearity of the latter, or the different symmetries characterizing each theory. However, in the non-relativistic limit, Coulomb and Newton forces share similarities in the properties we are focused on here.

The classical Lagrangian density is
\begin{eqnarray}
{\cal L} = \frac{1}{2} (D_{\mu}\phi)^* D^{\mu} \phi - m^2 |\phi|^2 - \frac{1}{4} F_{\mu \nu}F^{\mu \nu},
\end{eqnarray}
where $F_{\mu \nu} = \partial_{\mu} A_{\nu} - \partial_{\nu} A_{\mu}$ and $D_{\mu} = \partial_{\mu} - i e A_{\mu}$.

We define the conjugate momenta $\pi$ of the scalar field, and the EM vector potential $A_0,A_a (a=1,2,3)$  respectively as:
\begin{eqnarray}
\pi = \frac{\partial {\cal L}}{\partial \dot{\phi}} = \dot{\phi}^* \hspace{1cm}
p^0 = \frac{\partial {\cal L}}{\partial \dot{A}_0} = 0 \hspace{1cm}
E^a = \frac{\partial {\cal L}}{\partial \dot{A}_a} = F_{0a}.
\end{eqnarray}
The Hamiltonian is
\begin{eqnarray}
H = \int d^3x \left[|\pi|^2 + \partial_a \phi^* \partial^a \phi + m^2 |\phi|^2
+  \frac{1}{2}E^aE_a - \frac{1}{2}A_a(\nabla^2 A^a - \partial^a \partial_bA^b)
\right. \nonumber \\
\left.
-A_0 (\partial_aE^a - \hat{\varrho}) + J^a A_a + e^2 |\phi|^2A_aA^a\right]
\end{eqnarray}
where
\begin{eqnarray}
\varrho = ie (\phi^* \pi^* - \phi^* \pi^*)   \hspace{1.2cm}
J^a = ie (\phi \partial^a \phi^* - \phi^* \partial^a \phi)
\end{eqnarray}
are the charge density and the electric current respectively.

The system is characterized by the first class constraint (Gauss' law)
\begin{eqnarray}
\partial_aE^a - \varrho  = 0  \label{gauss}
\end{eqnarray}

The longitudinal components of $A^a$ are pure gauge (and can be taken for convenience to vanish) and the longitudinal components of $E_a$ are fixed by Gauss law.  Thus, the true degrees of freedom correspond to the transverse components ${}^T \! \!E^a$ of the electric field, the transverse components ${}^T\!\!A_a$  of  the magnetic potential and the  complex fields $\phi$ and $\pi$ corresponding to charged particles. The Hamiltonian expressed in terms of the true degrees of freedom is
\begin{eqnarray}
H = \int d^3x \left[|\pi|^2 + \partial_a \phi^* \partial^a \phi + m^2 |\phi|^2 + \int d\bfr d\bfr' \frac{\varrho({\bf r}) \varrho({\bf r'})}{4 \pi|{\bf r} - {\bf r'}|} \right.
\nonumber \\
\left. +   \frac{1}{2} {}^T\!\!E_a{}^T\!\!E^a - \frac{1}{2}{}^T\!\!A_a \nabla^2 {}^T\!\!A^a    + e^2 {}^T\!\!A_a {}^T\!\!A^a |\phi|^2\right]
\end{eqnarray}

Quantization proceeds in the standard way by expressing the field operators in terms of creation and annihilation operators $\hat{a}_{\bf k}$ and $\hat{a}^{\dagger}_{\bf k}$ for charged particles and $\hat{b}_{\bf k}$ and $\hat{b}^{\dagger}_{\bf k}$ for anti-particles.

\begin{eqnarray}
\hat{\phi}({\bf r}) = \int \frac{d^3k}{(2\pi)^3 \sqrt{2 \omega_{\bf k}}} \left[ \hat{a}_{\bf k} e^{i {\bf k}\cdot {\bf r}} + \hat{b}^{\dagger}_{\bf k} e^{- i {\bf k}\cdot {\bf r}}\right] \label{fieldq3a} \\
\hat{\pi}({\bf r}) = i \int \frac{d^3k}{(2\pi)^3 } \sqrt{\frac{ \omega_{\bf k}}{2}}\left[ - \hat{b}_{\bf k} e^{i {\bf k}\cdot {\bf r}} + \hat{a}^{\dagger}_{\bf k} e^{- i {\bf k}\cdot {\bf r}}\right]. \label{fieldq3b}
\end{eqnarray}
We now consider the non-relativistic limit for particles (rather than antiparticles). The fields $\hat{\psi}$ and $\hat{\psi}^{\dagger}$ are defined as in Eq. (\ref{psii}), and  the regularized charge density $\hat{\varrho}_{reg}({\bf r})$ is
 \begin{eqnarray}
\hat{\varrho}_{reg} ({\bf r}) = e \int d \bfr' \varsigma_{\sigma}({\bf r} - {\bf r'})\hat{\psi}^{\dagger}({\bf r}') \hat{\psi}({\bf r}')
\end{eqnarray}
where $\varsigma_{\sigma}({\bf r})$ is   the Gaussian function (\ref{gaussian}).

The field Hamiltonian becomes
\begin{eqnarray}
\hat{H} =  - \frac{1}{2m}\int d \bfr \hat{\psi}^{\dagger}({\bf r}) \nabla^2 \hat{\psi}({\bf r}) +   \int d \bfr d \bfr' \frac{\hat{\varrho}_{\sigma}({\bf r}) \hat{\varrho}_{\sigma}({\bf r'})}{4\pi|{\bf r} - {\bf r'}|}.
\end{eqnarray}
We then compute the Hamiltonian in the $N$-particle subspace
\begin{eqnarray}
\hat{H}_N = N m_{ren} \hat{1} + \sum_{i = 1}^N \frac{\hat{{\bf p}}^2_i}{2m} + \sum_{i \neq j} \frac{ e^2}{4 \pi |\hat{\bf r}_i - \hat{\bf r}_j|}, \label{hamqed}
\end{eqnarray}
where the renormalized mass  $m_{ren} = m + \delta m_{QED}$ includes a divergent term
\begin{eqnarray}
\delta m_{QED} = \frac{e^2}{4 \pi^{3/2} \sigma}.
\end{eqnarray}

For $N$ particles, at the limit $N \rightarrow \infty$,  the mean field theory approximation holds. We consider $N$-particle states of the form
\begin{eqnarray}
|\Psi\rangle = |\chi\rangle \otimes |\chi \rangle \ldots \otimes|\chi \rangle := \otimes_{i = 1}^N |\chi\rangle
\end{eqnarray}
   where $\chi$ is a single-particle wave-function. Then
\begin{eqnarray}
\lim_{N \rightarrow \infty} e^{-i \hat{H}t} \otimes_{i = 1}^N |\chi\rangle = \otimes_{i = 1}^N |\chi(t) \rangle
\end{eqnarray}
$\chi( t)$, a collective variable of the whole system under the mean field approximation, satisfies the Schr\"odinger-Coulomb  equation.
\begin{eqnarray}
i \frac{\partial}{\partial t} \chi({\bf r},t) = - \frac{1}{2m} \nabla^2 \chi({\bf r},t) + e^2 \int d \bfr' \chi({\bf r},t) \frac{|\chi({\bf r'},t)|^2}{4 \pi |{\bf r} - {\bf r'}|}  \label{hartree}
\end{eqnarray}
which is essentially the time-dependent version of Hartree's equation.

\bigskip

The QED case exemplifies our calculation for gravity. First, there is  no $N$-particle Schr\"odinger-Coulomb equation at the non-relativistic limit of QED. If the reasoning leading to the $N$-particle NS equations were applied to QED, we would obtain an equation of the form

\begin{eqnarray}
i \frac{\partial \psi}{\partial t}({\bf r}_1, \ldots, {\bf r}_N) = -\frac{1}{2m} \sum_i \nabla^2_i \psi({\bf r}_1, \ldots, {\bf r}_N)  \nonumber \\
+ \int  d {\bf X}' \frac{e^2}{|{\bf r}_i - {\bf X'}|} \rho_1({\bf X'}) \psi({\bf r}_1, \ldots, {\bf r}_N),
\end{eqnarray}
where
\begin{eqnarray}
\rho_1({\bf X}) = \sum_j \int d\bfr_1 \ldots d\bfr_N |\psi({\bf r}_1,\ldots, {\bf r}_N)|^2 \delta({\bf X} - {\bf r}_j)
\end{eqnarray}
is the `charge density' of the $N$ particles. This equation cannot account even for the most elementary results of quantum theory -- its analogue for one proton and one electron could not even predict the hydrogen-atom spectrum.

In contrast the standard evolution for  the $N$-particle wave function $\psi({\bf x}_1, \ldots, {\bf x}_n)$
\begin{eqnarray}
i \frac{\partial \psi}{\partial t} = \left(-\frac{1}{2m} \sum_i \nabla^2_i + \sum_{i \neq j} \frac{ e^2}{4 \pi |\hat{\bf r}_i - \hat{\bf r}_j|}\right) \psi,
\end{eqnarray}
follows directly from Eq. (\ref{hamqed}), modulo the mass term.

Second,  a ``semiclassical QED" approximation, corresponding to the equation $\partial_{\nu} F^{\nu \mu} = \langle \hat{j}^{\mu}\rangle$  is only meaningful at the level of the mean field theory with large number of particles.   Eq. (\ref{hartree}), viewed as a mean-field equation, applies in this regime.


\section{Conclusion}

We have given a summary of the main findings in the Introduction. Here we list the key points as conclusion:

\begin{enumerate}

\item {\em Coupling of classical gravity with quantum matter.}
The only viable theory for the description of matter degrees of freedom  is in terms of relativistic quantum fields. The coupling of classical gravity with quantum matter is meaningful  only under a mean field approximation for a large number of particles. The semiclassical Einstein equation operates under this condition. When fluctuations of quantum fields are included as source, the upgraded Einstein-Langevin equation describes the dynamics of the induced metric fluctuations. When passing to the non-relativistic limit one ought to describe quantum matter in terms of the  non-relativistic fields $\hat{\psi}(x), \hat{\psi}^{\dagger}(x)$ that correspond to annihilation and creation operators of particles respectively. Gravity couples to the mass-density which is an operator for a quantum system; assuming it be a c-number quantity leads one astray.

\item {\em Perils of single-particle wave function.}
The Newton-Schr\"odinger equation for the wave function of a single particle does not follow from general relativity and quantum field theory. Similarly, there is no  $N$-particle Newton-Schr\"odinger equation in gravity. When treating a system of N particles with large N, one can use an equation like the single-particle NS equation, but the wave function $\psi$ is a collective variable of the whole system of N particles under the mean-field approximation, not referring to a single particle.

\item  {\em No place for nonlinearity.}
There are severe obstacles to any non-linear Schr\"odinger equation for wave functions that define probabilities according to Born's rule.   This is not a specific problem of the NSE. Any theory involving a non-linear modification of Schr\"odinger's equation ought to explain how the probabilistic descriptions of quantum mechanics come about, since the most general transformation that preserves the probabilistic structure or quantum states is linear (at the level of density matrices) \cite{choi}. Non-linear Schrodinger-type equations such as the Hartree-Fock or the Gross-Pitaevski equations involve  wave functions $\Psi$ that are  collective variables for a many-body system, not single-particle quantum states. Theories based on NSEs entail unknown and hitherto ill-justified physics.

\end{enumerate}

\bigskip

\noindent {\bf Appendix}

\section{Derivation of the Hamiltonian for matter-gravity interaction at the weak field limit}
Here, we present the derivation of the Hamiltonian (\ref{hfin0}) from the action (\ref{Scov}) in the linearized-gravity approximation.

\subsection{The action}
We assume for the  spacetime manifold a spacelike foliation in the form $R \times \Sigma $ with time $t \in R$ and spatial coordinates $x^i$ on a spacelike surface $\Sigma$.  We denote the Riemannian metric on $\Sigma$ as $h_{ij}$ and the corresponding Ricci scalar as ${}^3R$. With this we perform a $3+1$ decomposition of the action   (\ref{Scov}) resulting in:
\begin{eqnarray}
S_{3+1}[h_{ij}, \phi, N, N^i] = \frac{1}{\kappa} \int dt d^3 x N \sqrt{h} \left[ K_{ij}K^{ij} -K^2 +{}^{(3)}R \right. \\ \nonumber
\left. + \frac{1}{2N^2}\dot{\phi}^2 - \frac{1}{2} (h^{ij} - \frac{N^iN^j}{N^2}) \nabla_i \phi \nabla_j \phi -\frac{1}{N^2} \dot{\phi} N^i\nabla_i\phi \right], \label{S3+1}
\end{eqnarray}
where $N$ is the lapse function, $N^i$ the shift vector, and
\begin{eqnarray}
K_{ij} = \frac{1}{2N} \left(\dot{h}_{ij} - \nabla_i N_j - \nabla_j N_i\right)
\end{eqnarray}
is the extrinsic curvature on $\Sigma$. The dot denotes taking the Lie
derivatives with respect to the vector field ${\partial}/{\partial t}$.

We consider perturbations around the Minkowski spacetime ($N = 1, N^i
= 0, h_{ij} = \delta_{ij}$) that are first-order with respect to
$\kappa$. That is, we write
\begin{eqnarray}
h_{ij} = \delta_{ij} + \kappa \gamma_{ij}, \hspace{1cm} N = 1 +
\kappa n, \hspace{1cm} N^i = \kappa n^i, \label{expand}
 \end{eqnarray}
 and we keep in Eq. (\ref{S3+1}) only terms up to first order in $\kappa$. We obtain
\begin{eqnarray}
S_{lin}[\gamma_{ij},\phi, n, n^i] =  \int dt d^3 x  \left(\frac{1}{2} \dot{\phi}^2 - \frac{1}{2}  \partial^i \phi \partial_i \phi - \frac{1}{2} m^2 \phi^2 \right) \nonumber \\
+ \kappa \int dt d^3x \left[ \frac{1}{4} (\dot{\gamma}_{ij} -2\partial_{(i} n_{j)})( \dot{\gamma}^{ij} - 2 \partial^{(i} n^{j)}) - \frac{1}{4} (\dot{\gamma} - 2\partial_i n^i)^2 \right.
\nonumber \\
\left.
- V[(\partial \gamma)^2] +n(\partial_i\partial_j \gamma - \partial^2\gamma)\right]  \nonumber \\
+ \frac{\kappa}{2} \int dt d^3x\left[(\frac{1}{2}\gamma - n)
\dot{\phi}^2 - 2n^i \dot{\phi}\partial_i \phi  + \gamma^{ij}
\partial_i \phi \partial_j \phi - (n+ \frac{1}{2} \gamma)
(\partial^i\phi \partial_i \phi + m^2 \phi^2)\right]. \hspace{1cm} \label{slin}
\end{eqnarray}
The indices in Eq. (\ref{slin}) are raised and lowered with the background 3-metric $\delta_{ij}$. We have defined $\gamma = \delta^{ij} \gamma_{ij}$. The ``potential" $V[(\partial \gamma)^2]$ corresponds to the second order terms in the expansion of $\sqrt{h} {}^3R$ with respect to $\gamma$; it will not be given, as it is not needed in the paper.

The first term in Eq. (\ref{slin}) is the action for a free scalar
field on Minkowski spacetime, the second term describes the self-dynamics
of the perturbations and the third term describes the
matter-gravity coupling.

\subsection{The Hamiltonian}

To obtain the Hamiltonian we perform the Legendre transform of the
Lagrangian density ${\cal L}_{lin}$ associated  to the action Eq.
(\ref{slin}). The conjugate momenta $\Pi^{ij}$ and $\pi$ of
$\gamma_{ij}$ and $\phi$ respectively are
\begin{eqnarray}
\Pi^{ij} &:=& \frac{\partial {\cal L}_{lin}}{\partial \dot{\gamma}_{ij}} = \frac{\kappa}{2} \left(\dot{\gamma}_{ij} -  \dot{\gamma}\delta^{ij} + \partial^in^j + \partial^j n^i -2 \partial_k n^k \delta^{ij}\right),\\
\pi &:=&  \frac{\partial {\cal L}_{lin}}{\partial \dot{\phi}} =
\dot{\phi} + \kappa \left[(\frac{1}{2} \gamma - n) \dot{\phi} - n^i
\partial_i \phi\right].
\end{eqnarray}
The conjugate momenta $\Pi_{n} = \partial {\cal
L}_{lin}/\partial{\dot{n}}$ and $\Pi^i_{
\overrightarrow{n}} = \partial {\cal L}_{lin}/\partial{\dot{n_i}}$
vanish identically. Thus, the equations $\Pi_{n} = 0$ and
$\Pi_{ \overrightarrow{n}}^i = 0$ define primary
constraints.

The Hamiltonian $H = \int d^3x (\Pi^{ij} \dot{\gamma}_{ij} + \pi \dot{\phi}
- {\cal L}_{lin})$  is
\begin{eqnarray}
H =  \int d^3 x \left[\left(\frac{\Pi^{ij}\Pi_{ij} - \frac{1}{2}
\Pi^2}{\kappa} + \kappa V[(\partial \gamma)^2]\right) +
\epsilon(\phi, \pi) \right. \nonumber \\
\left.
+ \frac{\kappa}{2}   \left[\gamma \epsilon(\phi, \pi) +  \gamma^{ij} \partial_i\phi \partial_j \phi - \gamma (\partial_k \phi \partial^k \phi +m^2 \phi^2)\right] \right. \nonumber \\
+ \left.   n \left[\partial^2\gamma - \partial_i \partial_j
\gamma^{ij} + \epsilon(\phi,\pi)\right] + n_i \left[-2
\partial_j \Pi^{ji} +  \kappa \mathfrak{p}^i(\pi, \phi) \right] \right],
\label{hlin}
\end{eqnarray}
where $\Pi = \Pi^{ij} \delta_{ij}$,
$\epsilon(\phi, \pi)$ is the energy density of the scalar field, Eq. (\ref{eden}), and
$\mathfrak{p}^i(\phi,\pi) = \pi \partial^i \phi$
is the momentum density.

\subsection{Constraints, Symmetries and Gauge-Fixing}

Eq. (\ref{hlin}) reveals the presence of secondary, first-class
constraints that arise from the usual scalar and vector constraints
of general relativity after linearization. The scalar constraint

\be
{\cal C} = \partial^2\gamma - \partial_i \partial_j \gamma^{ij} +
\epsilon = 0 \label{scalcon}
\ee
 generates the gauge transformations
\begin{eqnarray}
\delta \gamma_{ij} = 0, \hspace{0.5cm} \delta \Pi^{ij} = -
\partial^2\lambda \delta^{ij} + \partial^i \partial^{j}\lambda,
\hspace{0.5cm} \delta \phi = \lambda \frac{\delta H_0}{\delta \pi},
\hspace{0.5cm} \delta \pi = - \lambda  \frac{\delta H_0}{\delta \pi},
\label{gauge1}
\end{eqnarray}
where $H_0 = \int d^3x \epsilon$ is the field Hamiltonian at
Minkowski spacetime, and $\lambda$ is a scalar function on $\Sigma$. The
vector constraint

\be
{\cal C}^i := -2 \partial_j \Pi^{ji} +  \kappa
\mathfrak{p}^i = 0 \label{veccon}
\ee
 generates the gauge transformations
\begin{eqnarray}
\delta \gamma_{ij} = \partial_i \lambda_j + \partial_j \lambda_i,
\hspace{0.5cm} \delta \Pi^{ij} =0, \hspace{0.5cm} \delta \phi =
\kappa \lambda^i \partial_i \phi, \hspace{0.5cm} \delta \pi = \kappa
\partial_i(\lambda^i \pi) \label{gauge2}
\end{eqnarray}
where $\lambda^i$ is a vector-valued function on $\Sigma$.

The gauge transformations Eqs. (\ref{gauge1}---\ref{gauge2}) correspond to temporal and spatial reparameterizations of the free fields \cite{AHmasteq}.  The longitudinal part of the metric perturbation
${}^L\gamma_{ij}$ and the transverse trace ${}^T\Pi$ of the gravitational conjugate
momentum are pure gauge, reflecting the  freedom of space and time
reparameterization in the evolution of the matter degrees of freedom.

Next, we impose a gauge condition that preserves the Lorentz frame introduced by the foliation. We assume that
  ${}^L\gamma_{ij} = 0 $ and ${}^T\Pi = 0$.
In this gauge, the scalar constraint becomes the Poisson equation
$\partial^2\gamma =  - \epsilon$, which we   solve for $\gamma$ to
obtain
\begin{eqnarray}
\gamma(\bfr) =    \int d \bfr' \frac{\epsilon({\bf r'})}{4\pi |{\bf r} - {\bf
r'}|}. \label{gamma}
\end{eqnarray}
We also solve the vector constraint, in order to determine the
longitudinal part of $\Pi^{ij}$. We find
\begin{eqnarray}
{}^L\Pi^{ij}(\bfr) = i \int \frac{d^3k}{(2 \pi)^3} e^{-i {\bf k} \cdot
{\bf r}} [k_i \nu_j({\bf k}) + k_j\nu_i({\bf k})], \label{pil}
\end{eqnarray}
where
$\nu_i({\bf k}) = \frac{\kappa}{2} \left(\delta_{ij} -
\frac{k_ik_j}{2k^2}\right) \tilde{\mathfrak{p}}^j({\bf k})$;
   $\tilde{\mathfrak{p}}^i({\bf k})$ denotes the
Fourier transform of the momentum density $\mathfrak{p}^i$.

Thus the true physical degrees of freedom in the system correspond to the
transverse traceless components $\bar{\gamma}_{ij}$,
$\bar{\Pi}^{ij}$ of the metric perturbations and conjugate momenta,
and to the matter variables $\phi$ and $\pi$. The Hamiltonian
(\ref{hlin}) then  takes the form (\ref{hfin0}).

\section{Alternative derivation of the Hamiltonian (\ref{field1})}

Here, we sketch the derivation of the Hamiltonian (\ref{field1}) using a prescription of Dirac quantization, i.e., first quantizing and then solving the constraints. The derivation od the Hamiltonian (\ref{field1}) in the main text followed the reduced state space quantization, i.e., first solving the constraints and then quantizing.
The two methods are equivalent in the non-relativistic limit, thanks to  the simple form of the gravitational constraints take in this regime.

We start from a classical relativistic field interacting with gravity in the Newtonian approximation. The classical Hamiltonian for the scalar field is
    \begin{eqnarray}
    H = \frac{1}{2} \int d^3 x (1 - V_N) [  \pi^2 +    (\nabla \phi)^2 +   m^2 \phi^2], \label{ham3}
    \end{eqnarray}
    where $V_N$ is the Newtonian potential that satisfies Poisson's equation
    \begin{eqnarray}
    \nabla^2 V_N({\bf r}) = 4 \pi G \mu({\bf r}), \label{Poisson}
    \end{eqnarray}
 where $\mu$ is the mass density.

 The Hamiltonian (\ref{ham3}) leads to the Klein-Gordon equation
 \begin{eqnarray}
  \ddot{\phi} - \frac{\dot{V_N}}{1 - V_N} \dot{\phi} - (1 - V_N)^2 (\nabla^2 + m^2) \phi = 0,
 \end{eqnarray}
 or, to leading order in $V_N$
 \begin{eqnarray}
 \ddot{\phi} - \dot{V_N} \dot{\phi} - (1 - 2V_N) (\nabla^2 + m^2) \phi = 0
 \end{eqnarray}
 We quantize the system of equations by promoting the classical fields $\phi(x), \pi(x)$ to quantum operators (\ref{fieldq1}---\ref{fieldq2}) in the Hamiltonian.
  Then we pass to the Newtonian/non-relativistic limit as described by Eqs. (\ref{newt},  \ref{hamnr}).  The Hamiltonian becomes
 \begin{eqnarray}
 \hat{H} \simeq \hat{H}_0 - \int d \bfr V_N({\bf r}) \hat{\mu}_{\sigma}({\bf r})
 \end{eqnarray}

 Eq. (\ref{Poisson}) implies that the potential $V_N$ is a function of the mass-density operator $\hat{\mu}_{\sigma}({\bf r}) = m \hat{\psi}^{\dagger}({\bf r}) \hat{\psi}({\bf r})$, through the equation
\begin{eqnarray}
\hat{V}_N({\bf x}) = - G \int d \bfr' \frac{\hat{\mu}_{\sigma}({\bf r'})}{|{\bf r} - {\bf r'}|}.
\end{eqnarray}
{\em Thus $\hat{V}_N$ is also an operator.}

 Substituting $\hat{V}_N$ into the equation for the Hamiltonian, we obtain Eq. (\ref{field1}) modulo normal ordering.


\end{document}